\documentclass[preprint2]{aastex6}
\usepackage{amsmath,amstext}
\usepackage[figure,figure*]{hypcap}
\usepackage{newtxmath} 
\usepackage{subfigure}
\usepackage{hyperref}

\shorttitle{The Performance of Robo-AO at Kitt Peak}
\shortauthors{Jensen-Clem et al.}

\begin{document}

\title{The Performance of the Robo-AO Laser Guide Star Adaptive Optics System at the Kitt Peak 2.1-m Telescope}

\author{Rebecca Jensen-Clem\altaffilmark{1},  Dmitry A. Duev\altaffilmark{1}, Reed Riddle\altaffilmark{1}, Ma{\"i}ssa Salama\altaffilmark{2}, Christoph Baranec\altaffilmark{2}, Nicholas M. Law\altaffilmark{3}, S.~R.~Kulkarni\altaffilmark{1}
\& A. N. Ramprakash\altaffilmark{4}}

\altaffiltext{1}{Department of Astronomy, California Institute of Technology, 1200 E. California Blvd., Pasadena, CA 91101, USA}
\altaffiltext{2}{Institute for Astronomy, University of Hawai`i at M\={a}noa, Hilo, HI 96720-2700, USA}
\altaffiltext{3}{Department of Physics and Astronomy, University of North Carolina at Chapel Hill, Chapel Hill, NC 27599-3255, USA}
\altaffiltext{4}{Inter-University Centre for Astronomy \&\ Astrophysics,
Savitribai Phule Pune University Campus,
Pune 411 007, India}

\begin{abstract}
Robo-AO is an autonomous laser guide star adaptive optics system recently commissioned at the Kitt Peak 2.1-m telescope. Now operating every clear night, Robo-AO at the 2.1-m telescope is the first dedicated adaptive optics observatory. This paper presents the imaging performance of the adaptive optics system in its first eighteen months of operations. For a median seeing value of $1.31^{\prime\prime}$, the average Strehl ratio is 4\% in the $i^\prime$ band and 29\% in the J band. After post-processing, the contrast ratio under sub-arcsecond seeing for a $2\leq i^{\prime} \leq 16$ primary star is five and seven magnitudes at radial offsets of $0.5^{\prime\prime}$ and $1.0^{\prime\prime}$, respectively. The data processing and archiving pipelines run automatically at the end of each night. The first stage of the processing pipeline shifts and adds the data using techniques alternately optimized for stars with high and low SNRs. The second ``high contrast" stage of the pipeline is eponymously well suited to finding faint stellar companions.
\end{abstract}

\section{Introduction}

Adaptive optics (AO) systems correct wavefront aberrations introduced by the atmosphere and instrumental optics, restoring the resolution of a telescope to the diffraction limit. Laser guide stars (LGS) were developed in the 1980s to provide AO systems with bright, locatable wavefront reference sources, bringing fainter astrophysical objects into the purview of adaptive optics. Over half of all $>$8-m aperture telescopes are now equipped with an LGS AO system. The primary application of these AO instruments is for high angular resolution studies of interesting astronomical objects. As such minimizing the overhead has not been a major consideration in the overall design of AO systems on large telescopes.

Robo-AO is a robotic LGS AO system designed for maximum target throughput. Unlike LGS systems on large telescopes, it is based on an artificial star produced by Rayleigh scattering of a near UV laser. Robo-AO achieves high target throughput by minimizing overhead times to less than one minute per target. This is accomplished by three key design elements: 1) each step of the observation sequence is automated, allowing tasks that would be performed sequentially by a human operator to be performed in parallel and with minimal delay by the robotic system; 2) the $\lambda=355\,$nm Rayleigh scattering laser guide star is invisible to the human eye. As a result, while coordination with the U.S. Air Force Joint Space Operations Center (JSpOC) is still required to prevent illumination of sensitive space assets, no control measures are required by the Federal Aviation Administration; 3) Robo-AO employs an automated queue scheduler which chooses each new science target based on telescope slew times and approved lasing windows provided in advance by JSpOC. 

Robo-AO was first commissioned at the Palomar 1.5-m telescope in 2011, where it completed 19 science runs as a PI instrument from May 2012 through June 2015. Full details of the Robo-AO hardware and software can be found in \citet{baranec_bringing_2013}, \citet{baranec_high-efficiency_2014} and \citet{riddle_robo-ao_2014}. 

In 2012, the National Optical Astronomy Observatory (NOAO), following the recommendation of the Portfolio Committee which was chartered by the  the National Science Foundation (NSF),  decided to divest the Kitt Peak 2.1-m telescope.  In 2015, the Robo-AO team made a bid for the telescope and was selected to operate the telescope for three years.  Robo-AO was installed at the 2.1-m telescope in November, 2015; since then it has been operating nearly every clear night. As the first dedicated, automated adaptive optics facility, Robo-AO at Kitt Peak is well positioned to support the next generation of large-scale survey programs  that are focused on stellar and exoplanet astronomy (e.g.~K2, GAIA, CRTS, PTF, TESS and others), as well as AO follow up of interesting sources. Early science results including Robo-AO KP data can be found in \citet{adams_ultra-short-period_2017} and \citet{vanderburg_five_2016, vanderburg_two_2016}.

In this paper, we describe the performance of Robo-AO since commissioning. The paper is organized as follows: \S\ref{sec:SummaryInstrument} introduces the Robo-AO imaging systems; \S\ref{sec:reduction} provides an overview of our automatic data reduction pipelines; \S\ref{sec:seeing} shows the relationships between the weather conditions and the measured seeing; \S\ref{sec:analysis} presents the Strehl ratio and contrast curve statistics as well as the point spread function (PSF) morphology; \S\ref{sec:archive} describes our automated data archiving system; finally, \S\ref{sec:IRC} describes the newly installed near-IR camera.

\begin{table}[t]
 \centering
  \begin{tabular}{p{0.235\textwidth} p{0.2115\textwidth}}
  \hline \hline
Telescope     & Kitt Peak 2.1-m telescope \\
Science camera & Andor iXon DU-888 \\
EMCCD detector & E2V CCD201-20 \\
Read-noise (without EM gain) & 47 $e^{-}$ \\
EM gain, selectable & 300, 200, 100, 50, 25 \\
Effective read-noise & 0.16, 0.24, 0.48, 0.96, 1.9 $e^{-}$ \\
Full-frame-transfer readout & 8.6 frames per second \\
Detector format & 1024$^2$ 13-$\mu$m pixels\\
Field of view & 36\arcsec $\times$ 36\arcsec\\
Pixel scale & 35.1 milli-arcsec per pixel\\
Observing filters & $g^{\prime}$, $r^{\prime}$, $i^{\prime}$, $z^{\prime}$, lp600 \\
\hline
\end{tabular}
\caption{The specifications of the Robo-AO
optical detector at Kitt Peak.}
\label{tab:EMCCD}
\end{table}

\section{Summary of the Robo-AO Imaging System}
\label{sec:SummaryInstrument}

The Robo-AO imaging system includes two optical relays, each using a pair of off-axis parabolic mirrors. The first relay images the telescope pupil onto a 140-actuator Boston Micromachines micro-electro-mechanical-systems (MEMS) deformable mirror used for wavefront correction. A dichroic then reflects the UV light to an 11$\times$11 Shack-Hartmann wavefront sensor. The second optical relay includes a fast tip-tilt correcting mirror and an atmospheric dispersion corrector (ADC; here, two rotating prisms\footnote{From the commissioning of Robo-AO at Kitt Peak in November of 2015 through February of 2017, the right ascension (RA) axis of the 2.1-m telescope suffered from a $\sim 3.7\,$Hz jitter (see \S\ref{sec:strehl} and \S\ref{apend:jitter}) that caused a slight elongation of the stellar PSFs. As a result, the ADCs were not correctly calibrated until an upgrade to the telescope control system removed the jitter.}) located at a reimaged pupil. The output of the second relay is an F/41 beam that is intercepted by a dichroic mirror, which reflects the $\lambda<950\,$nm portion of the converging beam to the visible wavelength filter wheel and EMCCD detector (see Table \ref{tab:EMCCD}). The filter wheel includes  $g^{\prime}$, $r^{\prime}$, $i^{\prime}$, and $z^{\prime}$ filters, as well as a long-pass ``lp600" filter cutting on at 600\,nm and extending beyond the red limit of the EMCCD (see Figure~1 in \citealt{baranec_high-efficiency_2014}). The dichroic transmits the longer wavelength light to the near-infrared (NIR) instrument port (see \S\ref{sec:IRC}).

Robo-AO was originally designed for simultaneous optical and NIR operations, such that deep science integrations could be obtained in one band while the image displacement could be measured in the other and corrected with the fast tip-tilt mirror. In February of 2017, we achieved first light with a science-grade novel infrared array, a brief summary of which appears in \S\ref{sec:IRC}\footnote{A detailed analysis of the operation of this camera, its imaging performance, and its incorporation into an active tip-tilt control loop will be reported elsewhere.}. In this paper, we consider the imaging performance of Robo-AO using the optical imaging camera only. In lieu of an active tip-tilt correction, the EMCCD is run at a framerate of $8.6\,$Hz to allow for post-facto image registration followed by stacking (see \S\ref{sec:reduction}).

\section{Data Reduction Pipelines}
\label{sec:reduction}

\subsection{Overview}

Image registration and stacking (see \S\ref{sec:SummaryInstrument}) is accomplished automatically by the ``bright star'' and ``faint star'' pipelines, which are optimized for high and low signal-to-noise (SNR) targets, respectively. The data are then processed by the ``high contrast'' pipeline to maximize the sensitivity to faint companions. These pipelines are described in detail below.

\subsection{Image Registration Pipelines}
\label{sec:ImageRegistrationPipeline}
All observations are initially processed by the ``bright star'' pipeline. This pipeline generates a windowed datacube centered on an automatically selected guide star. The windowed region is bi-cubically up-sampled and cross correlated with the theoretical point spread function to give the center coordinates of the guide star's PSF in each frame. The full-frame, unprocessed images are then calibrated using the nightly darks and dome flats. Finally, the calibrated full frames are aligned using the center coordinates identified by the up-sampled, windowed frames, and co-added via the Drizzle algorithm \citep{fruchter_drizzle:_2002}. These steps are described in detail in \citet{law_twelve_2014}. 

After an observation has been processed by the ``bright star'' pipeline, the core of the brightest star in the frame is fit by a 2D Moffat function. If the full width at half maximum (FWHM) of the function fit to the core is  $<\lambda/$D, indicating that the stellar centroiding step has failed, the observation is re-processed by the ``faint star'' pipeline to improve the SNR in the final science image. 

\begin{figure}[htp]
  \centering
  \subfigure[Bright star, bright star pipeline]{\includegraphics[width=3.9cm]{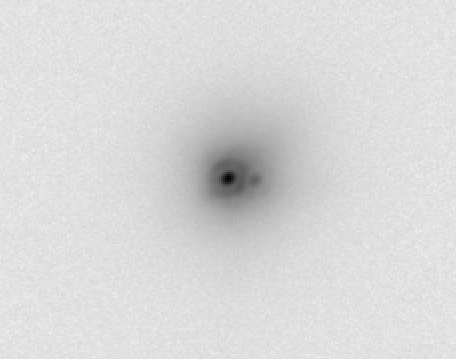}}\quad
  \subfigure[Bright star, faint star pipeline]{\includegraphics[width=3.9cm]{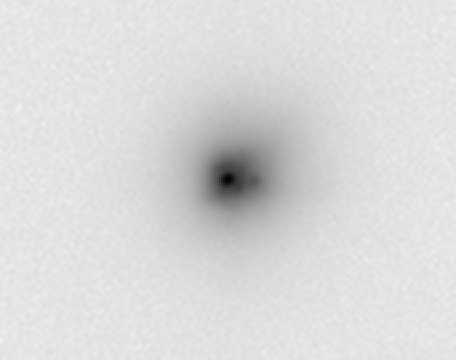}}
  \subfigure[Faint star, bright star pipeline]{\includegraphics[width=3.9cm]{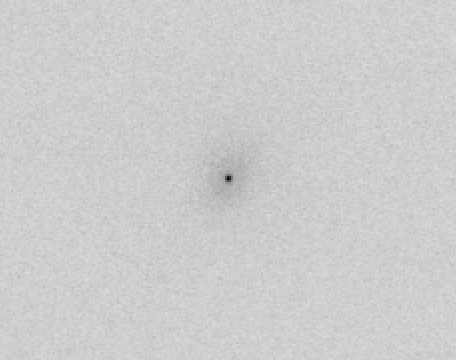}}\quad
  \subfigure[Faint star, faint star pipeline]{\includegraphics[width=3.9cm]{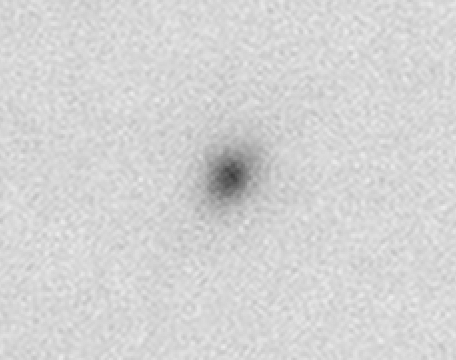}}
\caption{The bright star pipeline (a) produces a superior Strehl ratio for the V$=8.84$ double star HIP55872 compared with (b) the faint star pipeline. For the V$=15.9$ star 2MASSJ1701+2621, however, the bright star pipeline (c) fails to correctly center the PSF, leading to an erroneously bright pixel in the center. The faint pipeline (d) successfully shifts and adds this observation. }
\label{fig:pipelines}
\end{figure}

The individual frames for a given observation are summed to create a master, dark and flat corrected reference image. This frame is then high pass filtered and windowed about the guide star. Each raw short exposure frame is then dark and flat corrected, filtered, and windowed. These individual frames are registered to the master reference frame using the  \texttt{Image Registration for Astronomy} python package written by Adam Ginsburg\footnote{\url{https://github.com/keflavich/image_registration}}. The package finds the offset between the individual and reference frames using DFT up-sampling and registers the images with FFT-based sub-pixel image shifts. Figure \ref{fig:pipelines} illustrates the strengths and weaknesses of the bright and faint star pipelines. 

\begin{figure*}[htp]
    \centering
    \includegraphics[width=\textwidth]{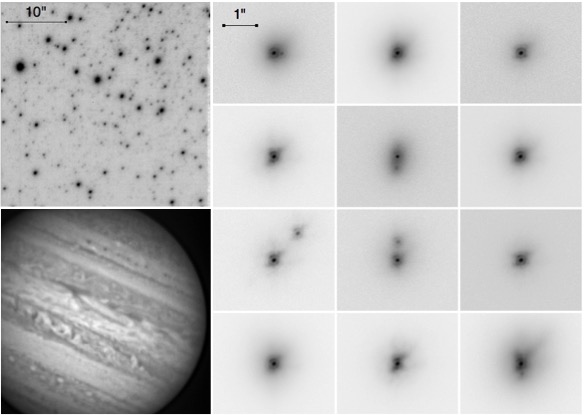}
    \caption{Examples of Robo-AO $i^{\prime}-$band images at Kitt Peak (square root scaling) The full-frame ($36^{\prime\prime}\times36^{\prime\prime}$) images on the left are    the globular cluster Messier 5 (top) and Jupiter (bottom). The images on the right are examples of bright single stars and stellar binaries with a range of separations and contrasts. } 
    \label{fig:allpictures}
\end{figure*}

These automatic pipelines have reduced thousands of Robo-AO observations since the instrument was commissioned in November of 2015. Figure \ref{fig:allpictures} shows a collage of representative observations. 

\subsection{High Contrast Pipeline}
\label{sec:HighContrastPipeline}
For science programs that aim to identify point sources at small angular separations from known stars further processing is needed. Our ``high contrast imaging'' pipeline generates a $3.5^{\prime \prime}$ frame windowed about the star of interest in the final science frame. A high pass filter is applied to the windowed frame to reduce the contribution of the stellar halo. To whiten correlated speckle noise at small angular separations from the target star we subtract a synthetic PSF generated by Karhunen-Lo\`{e}ve Image Processing (KLIP). The KLIP algorithm is based on the method of Principal Component Analysis \citep{soummer_detection_2012}. The PSF diversity needed to create this synthetic image is provided by a reference library of Robo-AO observations -- a technique called Reference star Differential Imaging \citep[RDI;][]{lafreniere_hst/nicmos_2009}. We note that the angular differential imaging approach \citep{marois_angular_2006} is not possible here because the 2.1-m telescope is an equatorial mount telescope. Our pipeline uses the \texttt{Vortex Image Processing (VIP)} package \citep{gomez_gonzalez_VIP}.

The full reference PSF library consists of several thousand $3.5^{\prime \prime}$ square high pass filtered frames that have been visually vetted to reject fields with more than one point source. The PSF library is updated on a nightly basis to ensure that each object's reduction has the opportunity to include frames from the same night. Each frame in the full library is cross correlated with the windowed and filtered science frame of interest. The five frames with the highest cross correlation form the sub-library provided to KLIP. We then adopt only the first principal component (PC) as our synthetic PSF, as including more PCs provides no additional noise reduction on average. A future version of the pipeline will choose the number of PCs automatically for each observation based on SNR maximization. 

\begin{figure}[htp]
  \centering
  \subfigure[PSF after standard pipeline reduction]{\includegraphics[width=2.62cm]{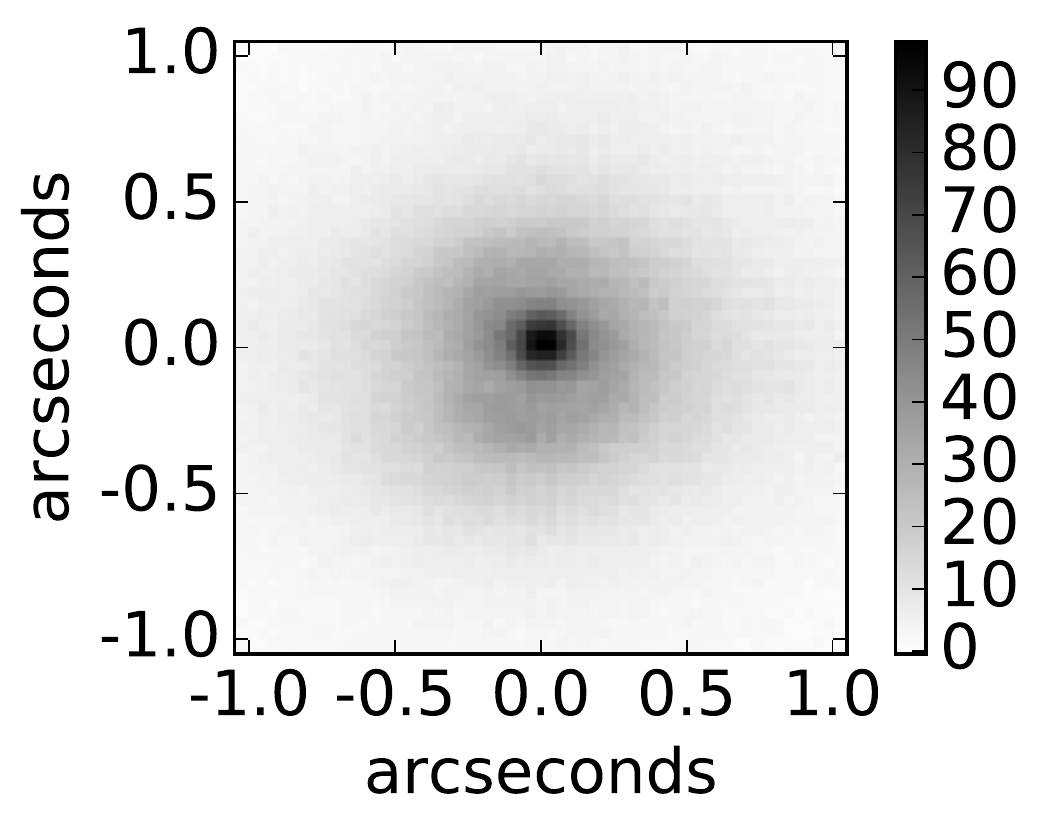}}\quad
  \subfigure[PSF in (a) after high pass filtering]{\includegraphics[width=2.62cm]{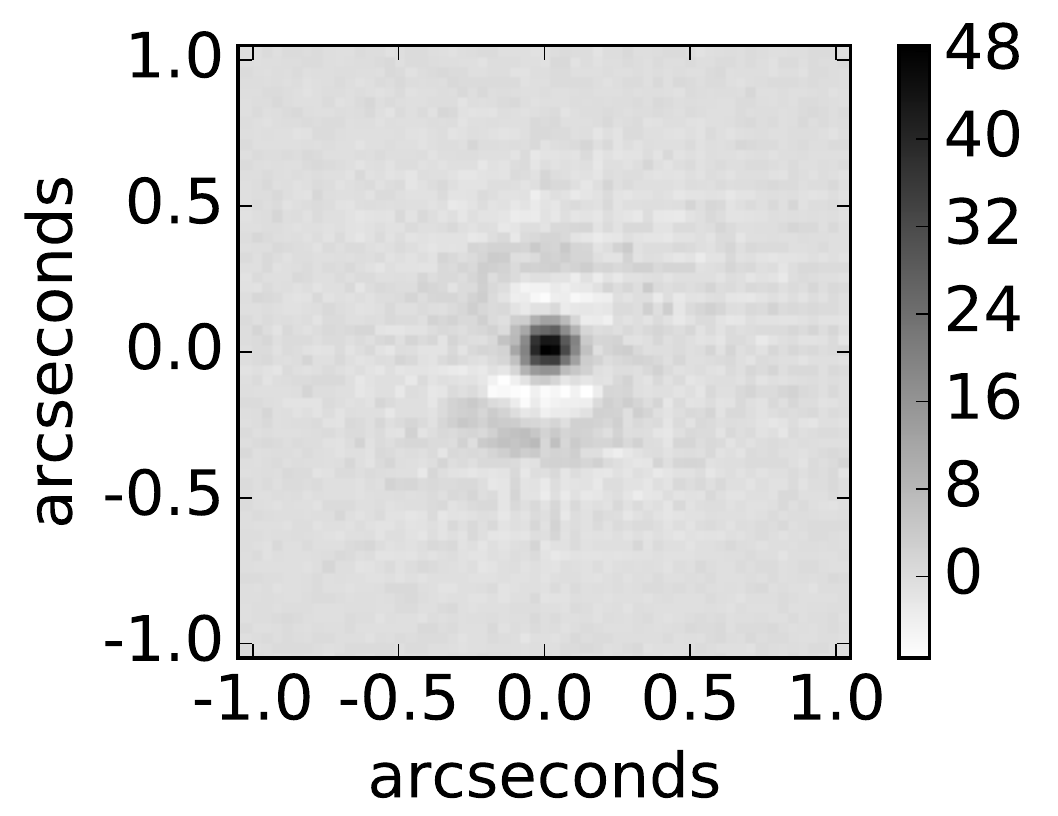}}\quad
  \subfigure[PSF in (b) after RDI+KLIP reduction]{\includegraphics[width=2.62cm]{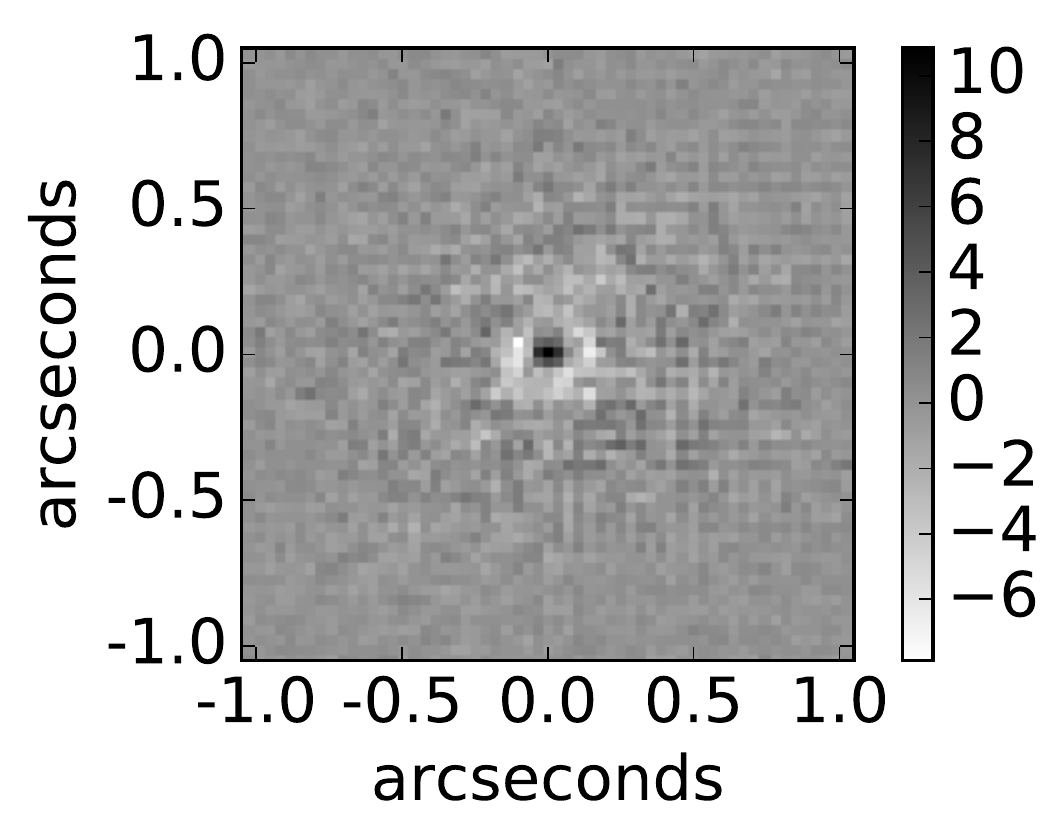}}
  \subfigure[The dashed, dot-dashed, and solid contrast curves correspond to the PSFs shown in (a), (b), and (c), respectively.]{\includegraphics[width=0.45\textwidth]{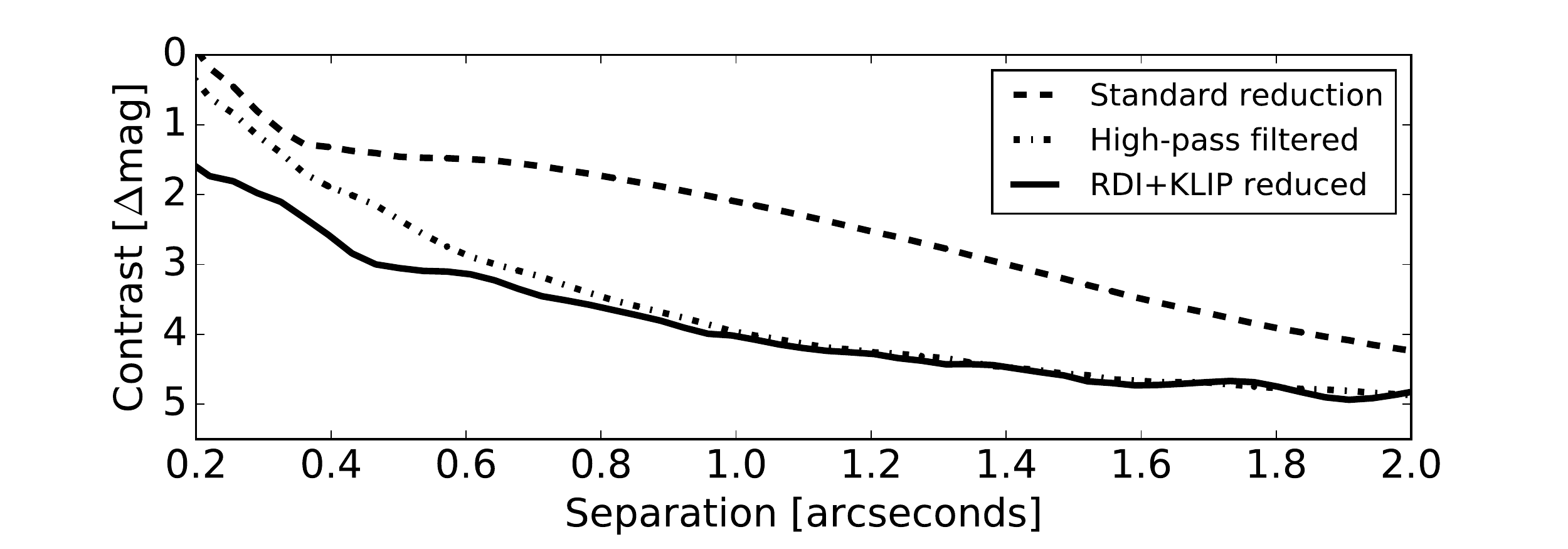}}
\caption{An example of the reduction steps in the ``high contrast'' pipeline for a $z^{\prime}$ observation of the star EPIC228859428. }
\label{fig:reduction}
\end{figure}

Figure \ref{fig:reduction} shows an example of a PSF reduced by the standard data pipeline (panel \texttt{a}), then high pass filtered (panel \texttt{b}), and finally processed with RDI-KLIP (panel \texttt{c}). 
After a science frame has been fully reduced we use  \texttt{VIP} to produce a contrast curve that is properly corrected for small sample statistics and algorithmic throughput losses. The corresponding contrast curves for the three panels are shown in panel \texttt{d}.  

Given that over two hundred new targets are observed during a clear night of Robo-AO observations the reference library is rapidly expanding and increasingly includes PSFs affected by a very wide range of environmental conditions. Hence, speckle noise in a past observation can be further reduced by a fresh RDI-KLIP reduction if the data is more correlated with later PSFs. Clearly a new reduction will benefit from the advantage of the larger reference library. 

\newpage
\section{Site Performance}
\label{sec:seeing}

\subsection{Site Geography}

Kitt Peak is located $56\,$miles southwest of Tucson, Arizona, at an elevation of $6800\,$feet. The 2.1-m telescope is situated $0.4\,$miles to the south of the peak's highest point (the location of the Mayall 4-m telescope). The WIYN 3.5-m and 0.9-m telescopes are respectively $700\,$ft and $400\,$ft to the west of the 2.1-m telescope and at approximately the same elevation. There are no structures at equal or greater elevations to the east of the telescope, and the terrain is relatively flat beyond Kitt Peak in that direction. The $7730\,$ft Baboquivari Peak is $12\,$miles directly south of the telescope. 

\subsection{Seeing Measurement}
\label{sec:SeeingMeasurement}

Before the start of each science observation, a $10\,$s seeing observation is taken with the AO correction off. During this period the wavefront sensor camera acquires a background image.  These seeing observations are dark and flat calibrated and summed without any registration of the individual exposures. The seeing is defined as the FWHM of a two-dimensional Gaussian function fit to this summed frame. Starting in January of 2017, a $90\,$s seeing observation was obtained each hour. Specifically, the Robo-AO queue schedules an observation of a bright ($V<8$) star within $10^{\circ}$ of zenith to refocus the telescope and measure the seeing. As of this writing, there is no significant difference between these ``long'' and ``short'' seeing observations. Here on we proceed with the assumption that the $10\,$s seeing measurements are representative of the long-exposure seeing. 

\begin{figure*}[htp]
    \centering
    \includegraphics[width=\textwidth]{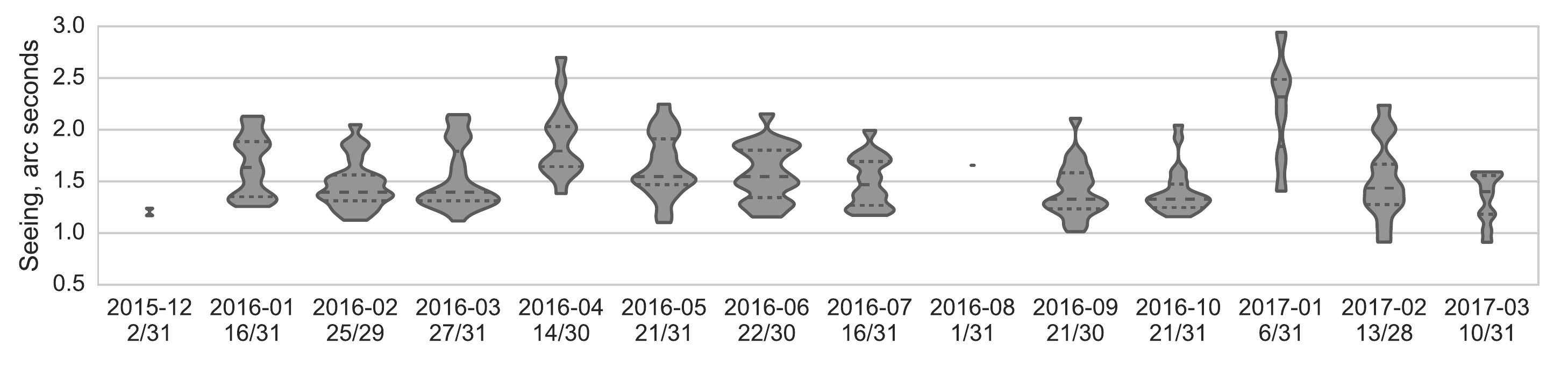}
    \caption{Seasonal fiducial ($\lambda=500\,$nm; see \S\ref{sec:SeeingMeasurement}) seeing measurements. Nightly median values were used to fit a monthly distribution. The fraction of nights with seeing data for each month is shown. The quartile values and the actual measured range are shown. 
    } 
    \label{fig:seeing_violin}
\end{figure*}

\begin{figure}[h!]
    \centering
    \includegraphics[width=0.45\textwidth]{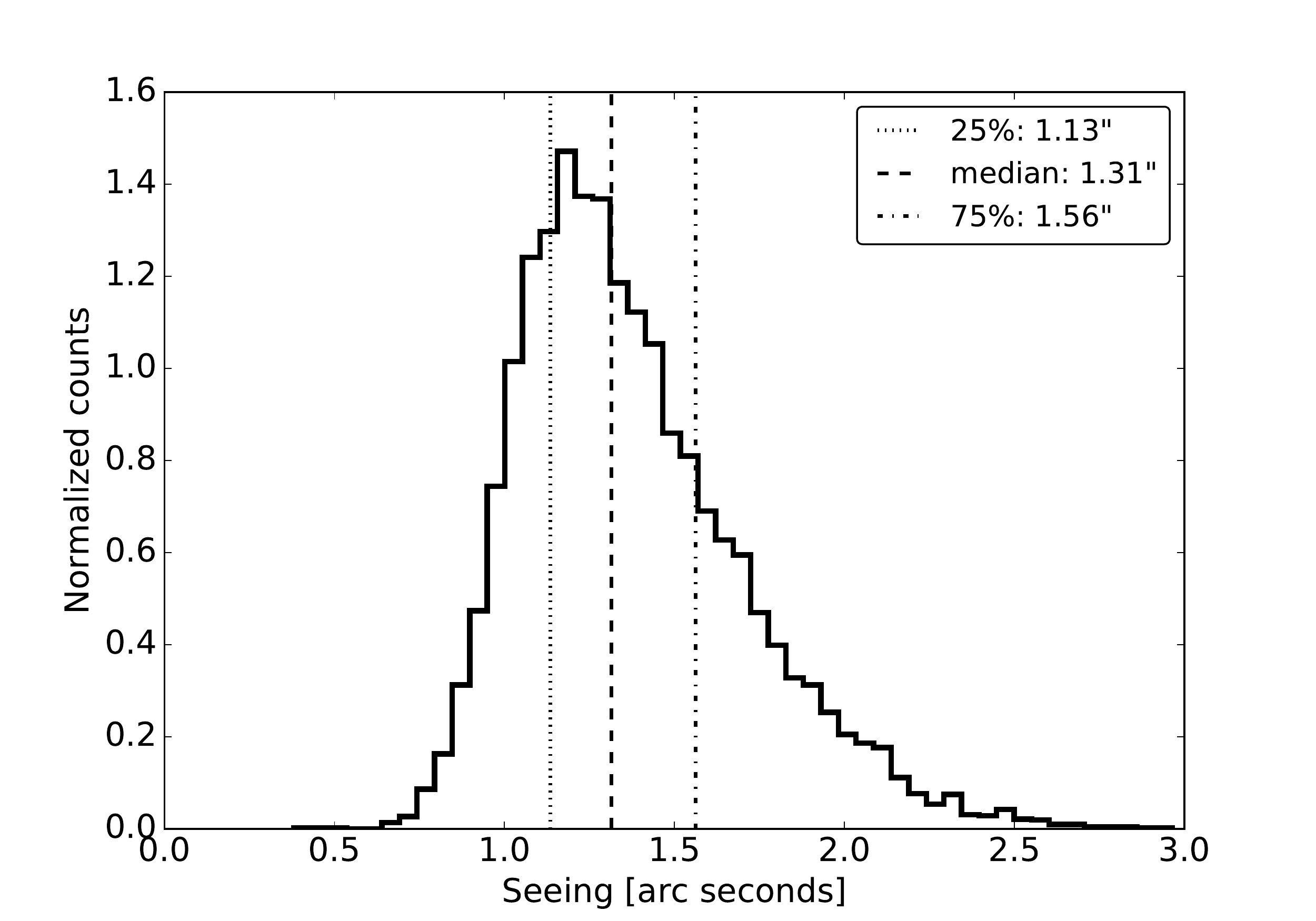}
    \caption{A histogram of the seeing measurements (all referenced to a wavelength $\lambda=500\,$nm) from December 2015 to March 2017. A zenith distance dependent correction has been applied. The 25th, 50th, and 75th percentile seeing values are indicated by the vertical lines. }
    \label{fig:seeing}
\end{figure}

We display a histogram of these fiducial seeing values in Figure~\ref{fig:seeing}.  Figure~\ref{fig:seeing_violin} displays the seeing as a function of the seasons. The seeing values measured in a given wavelength are scaled to a fiduciary wavelength of 500\,nm by the scaling law $\text{seeing}_{500\,\text{nm}} = \text{seeing}_\lambda \times (\lambda/\text{500}\,\text{nm})^{1/5}$.

\subsection{Seeing Contributions}

\begin{figure}[h!]
    \centering
    \includegraphics[width=0.45\textwidth]{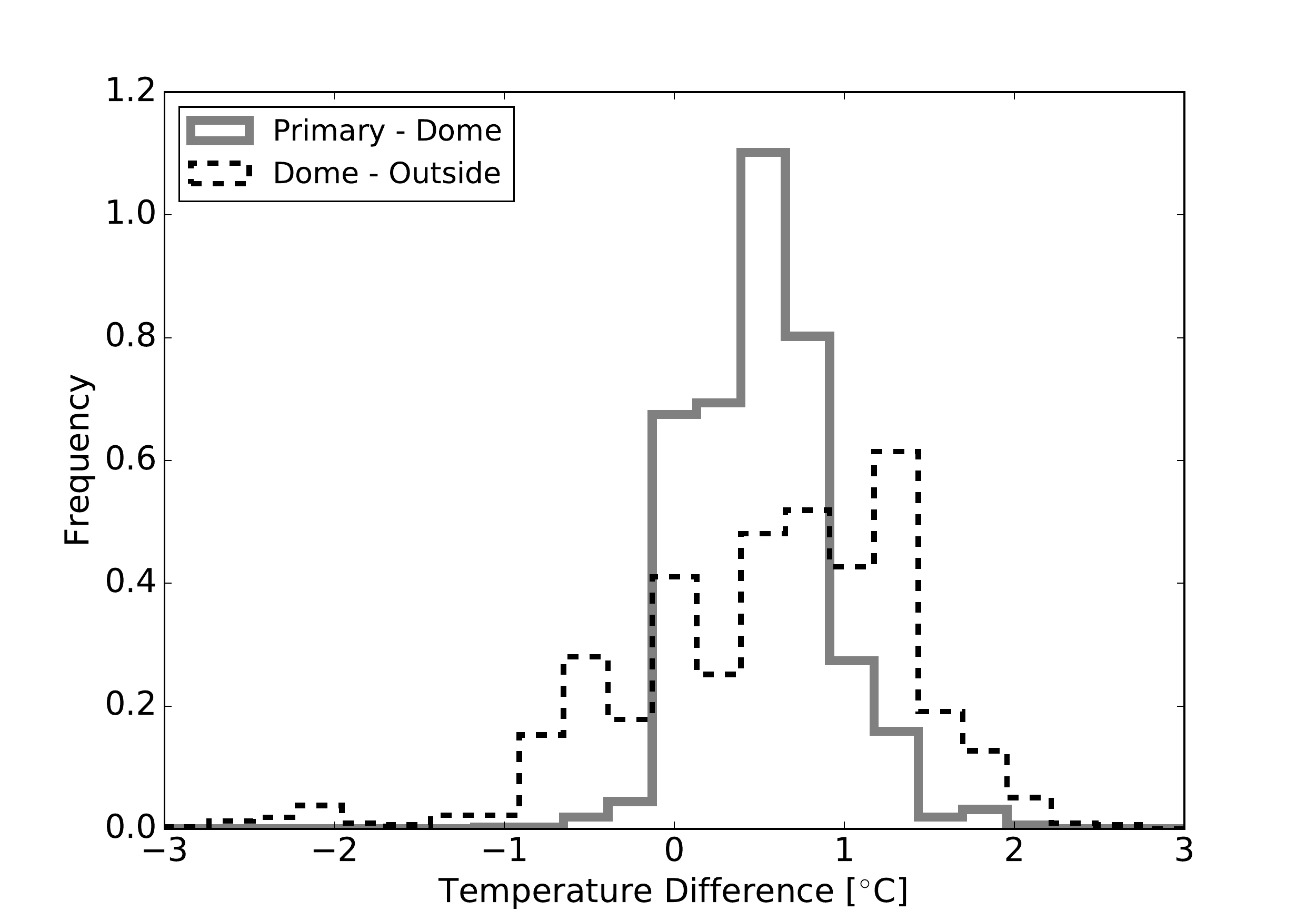}
    \caption{Histograms of the difference between the primary mirror and dome temperatures (solid) and the dome temperature minus the outside air temperature (dashed).}
    \label{fig:temp_hist}
\end{figure}

We note that our median seeing of $1.31^{\prime\prime}$ differs from the median seeing of $0.8^{\prime\prime}$ reported by the adjacent WIYN telescope\footnote{\url{ https://www.noao.edu/wiyn/aowiyn/}}. One possible explanation for this discrepancy is that  the WIYN was built in 1994 with careful attention paid to dome ventilation and telescope thermal inertia. In contrast, the 2.1-m telescope saw first light in 1964 before such considerations were fully appreciated. Figure \ref{fig:temp_hist} demonstrates the challenging thermal conditions at the 2.1-m telescope: during the majority of Robo-AO observations, the mirror is warmer than the ambient dome temperature which in turn is warmer than the outside air. The experience of other observatories indicate that improvements to dome thermalization can significantly improve the measured seeing  \citep[e.g.][]{bauman_dome_2014}.  

\begin{figure}[h!]
    \centering
    \includegraphics[width=0.45\textwidth]{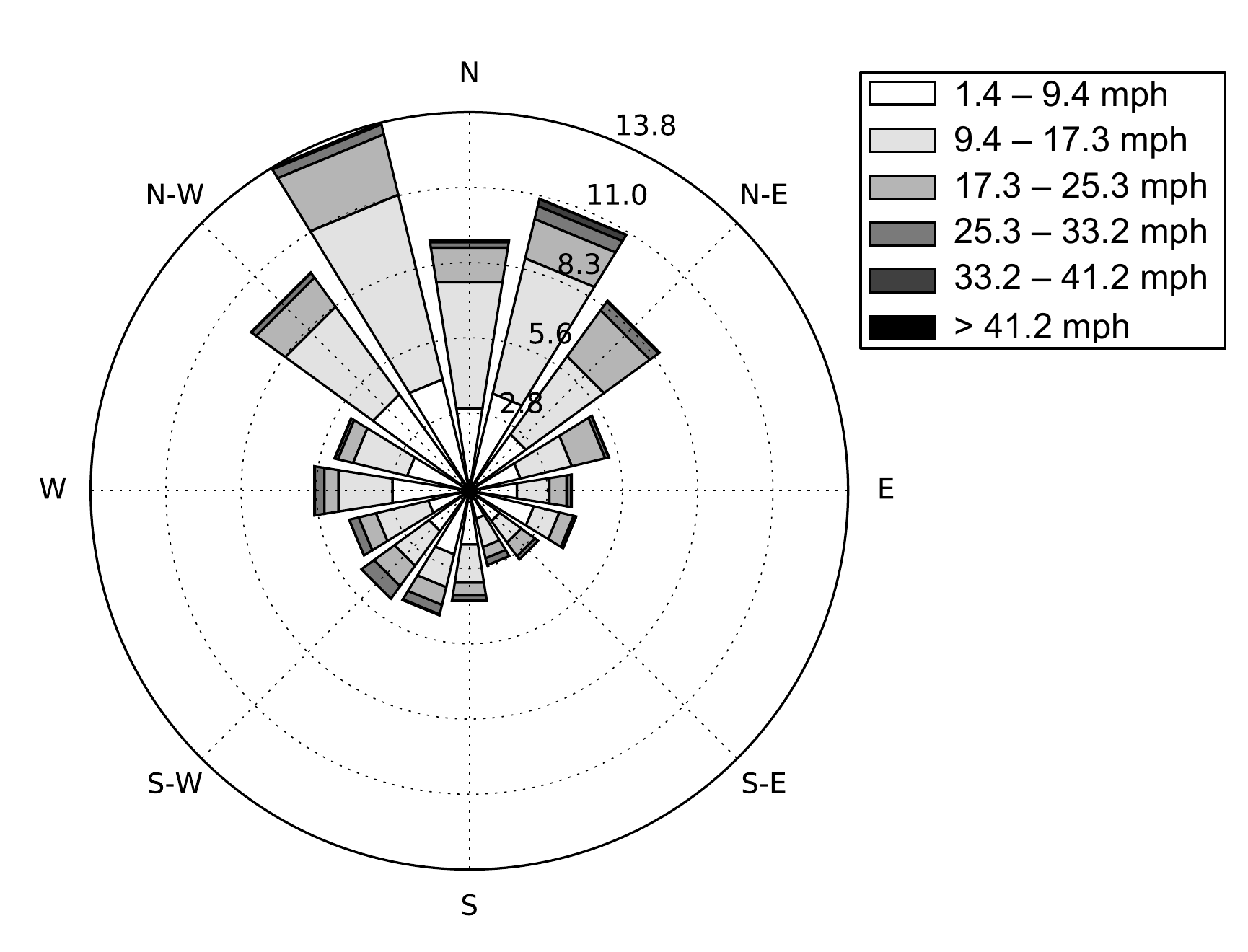}
    \caption{A ``wind rose" showing a stacked polar histogram of wind speeds and directions from December 2015 through June 2016. The wind most frequently blows from the NW, N, and NE, which correspond to the more mountainous region towards the direction of the Mayall 4-m telescope. These also tend to be the direction of the high wind speeds while slower wind speeds most often come from the south, where the terrain is less mountainous.}
    \label{fig:windrose}
\end{figure}

Another possible cause of the comparatively poor seeing at the 2.1-m telescope is perhaps a more turbulent ground layer. Figure \ref{fig:windrose} shows a ``wind rose,'' or the frequency of wind speeds originating from different directions, for December 2015 through June 2016. We find that during this period the wind most commonly blows from the NNW, or the direction of the higher elevation Mayall 4-m telescope, and rarely from the SE where the terrain is less mountainous. The highest winds ($>40\,$mph) come from the north while the south has the largest fraction of low wind speeds (the wind speeds originating from within $20^{\circ}$ of due south are under $10\,$mph $73\%$ of the time). 

Despite these terrain variations, the seeing is not significantly correlated with the wind direction. The wind speed, however, degrades the seeing by several tenths of an arcsecond for winds over $20\,$mph (the dome closes for winds over $40\,$mph). 

\begin{figure}[h!]
    \centering
    \includegraphics[width=0.45\textwidth]{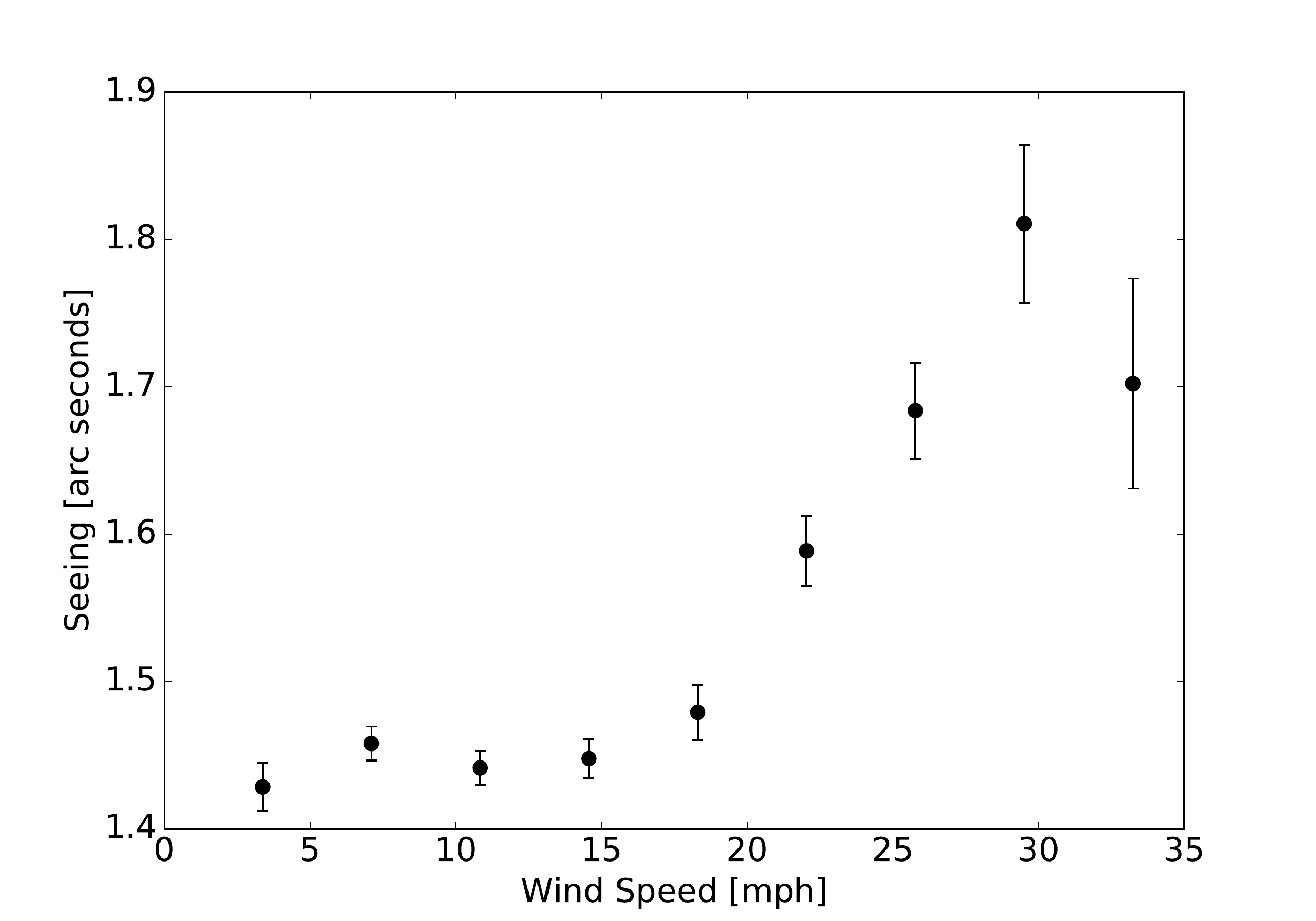}
    \caption{The mean binned seeing versus the wind speed for December 2015 through June 2016. The error bars are the standard deviation of the seeing values in a given wind speed bin divided by the square root of the number of seeing measurements in the bin. For wind speeds over $20\,$mph, the seeing is degraded by up to $0.3^{\prime \prime}$.}
    \label{fig:seeing_speed}
\end{figure}

Figure \ref{fig:seeing_speed} plots the seeing versus the wind speed, demonstrating that poorer seeing is correlated with higher wind speeds\footnote{The mean binned seeing measurements in Figure \ref{fig:seeing_speed} are larger than the median of all Robo-AO KP seeing measurements (Figure \ref{fig:seeing}) due to binning effects and the difference between the mean and median of the asymmetric distribution of seeing measurements.}. We note that the wind monitor became nonfunctional after June of 2016, and hence further study of the relationship between the seeing and the wind speed will occur after a new wind monitor is in place.

\section{Adaptive Optics Performance}
\label{sec:analysis}

\subsection{Strehl Ratio}
\label{sec:strehl}

The goal of an adaptive optics system is to bring the observed PSF closer to its theoretical diffraction-limited shape; hence, an important measure of the AO system's performance is the ratio between the peak intensity of an observed PSF and that of the telescope's theoretical PSF -- the Strehl ratio.  As the AO performance improves, the Strehl ratio increases.

We calculate the Strehl ratio by 1) generating a monochromatic diffraction-limited PSF by Fourier transforming an oversampled image of the pupil, 2) combining several monochromatic PSFs to create a PSF representative of the desired bandpass, 3) re-sampling the polychromatic PSF to match our 0.0175$^{\prime\prime}/$pixel platescale of the up-sampled ``drizzled'' frames, 4) obtaining the ``Strehl factor," or the ratio of the peak intensity to the sum of the intensity in a $3^{\prime\prime}$ square box, and 5) calculating the Strehl ratio by repeating step 4 for the observed image and dividing by the Strehl factor. These steps are described in detail in \citet{salama_robo-ao_2016}. 

Once Robo-AO began regular observations  at the 2.1-m telescope, we noticed that the achieved Strehl ratios were noticeably smaller than those that were achieved (for similar seeing values) at the Palomar 1.5-m telescope.  A number of exercises were undertaken to determine possible causes for this degradation. Eventually, we determined that the Telescope Control System (TCS) was the main contributing factor. In Appendix \ref{apend:jitter} we discuss the problem in detail. The mitigation consisted of upgrading the TCS (completed February 2017). Below, and for the rest of the paper, we discuss the instrument performance since the TCS upgrade. 

\begin{figure}[htp]
  \centering
 {\includegraphics[width=0.45\textwidth]{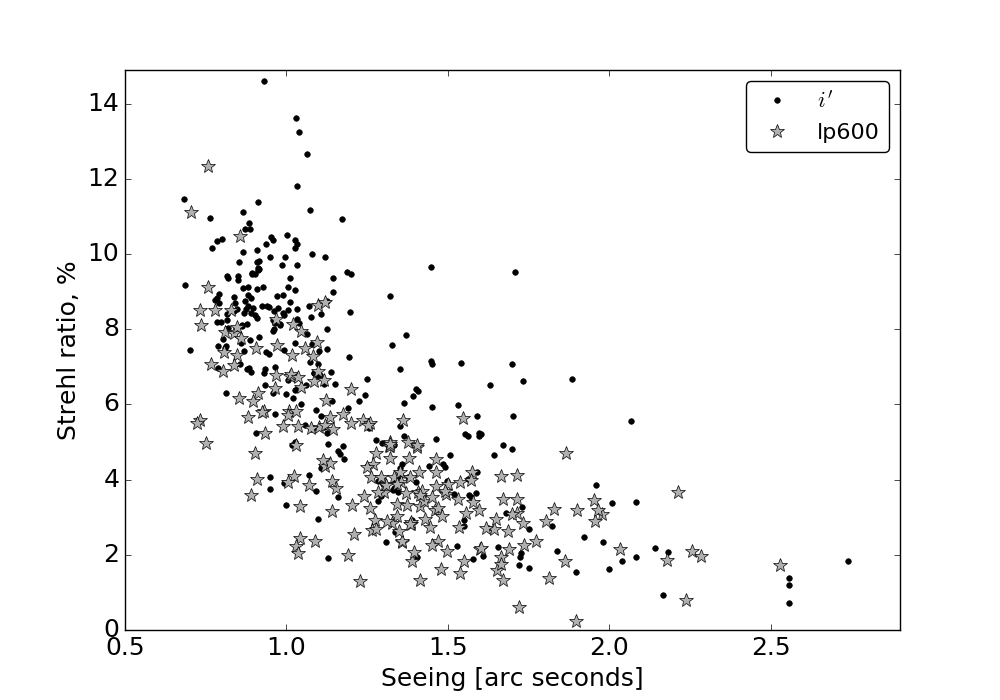}}
 \caption{The Strehl ratio versus the measured seeing values for 21 February 2017 through 10 March 2017 in the $i^{\prime}$ and lp600 filters.}
\label{fig:strehl_seeing}
\end{figure}

Figure \ref{fig:strehl_seeing} plots the Strehl ratio versus the measured seeing for the $i^{\prime}$ and lp600 filters. It is clear that the delivered Strehl ratio  drops off quickly as the seeing increases -- while Robo-AO achieves $>10\%$ Strehl ratio when the seeing is $<1.0^{\prime\prime}$, a $0.25^{\prime\prime}$ seeing increase halves the Strehl ratio. 

In Table \ref{tab:error}  we present a detailed error budget under different seeing conditions. This error budget was originally developed by R. Dekany (private communication), and was validated against the on-sky performance of laser AO systems on the Keck telescope, the Hale telescope and the Palomar 1.5-m telescope \citep{2012SPIE.8447E..04B}. Since we lacked   turbulence profile(s) for the 2.1-m telescope site we adopt a mean $C_{n}^{2}(h)$ profile from a MASS-DIMM atmospheric turbulence monitor collected over a year's baseline at Palomar and scaled to the seeing at Kitt Peak.

High-order errors are added in quadrature and are dominated by Focal Anisoplanatism (which is an error arising from the finite altitude of the Rayleigh laser guide star resulting in imperfect atmospheric sampling). We estimate one-axis tip-tilt errors as being dominated by bandwidth error for magnitudes greater than 13. As noted in \S\ref{sec:SummaryInstrument} we did not use the built-in tip-tilt facility but instead resorted to shift and add. We approximate the error resulting from this approach as follows.  We assume a standard $-3$db rejection frequency matching the full-frame rate of the science camera to approximate bandwidth error. The tip-tilt errors are then converted to an equivalent wavefront error and summed in quadrature with the high-order errors. Other high-order and tip-tilt errors include chromatic, scintillation, aliasing, calibration and digitization errors. 

Strehl ratios are calculated using the Mar\'{e}chal approximation. The full-widths at half-maximum (FWHM) are calculated from PSF models assuming the residual diffraction-limited, concentrated light, residual seeing, and scattered light halos are proportional to the phase variance of the residual errors. These models have shown accuracy of a few percent for Strehl ratios as low as 4\% \citep{2006ApJ...647.1517S}.
Figure \ref{fig:strehl_seeing} demonstrates Robo-AO's ability to approach the predicted Strehl ratio of $14\%$ in sub-arcsecond seeing conditions.  

\begin{table*}[htp]
 \centering
 \caption{The Robo-AO Error Budget}

  \begin{tabular}{p{0.3\textwidth} p{0.125\textwidth} p{0.125\textwidth} p{0.125\textwidth} p{0.125\textwidth}}
  \hline \hline

\textbf{Percentile Seeing} & & \textbf{25\%} & \textbf{50\%} & \textbf{75\%} \\
Seeing at Zenith & $1.00^{\prime\prime}$ & $1.13^{\prime\prime}$ & $1.31^{\prime\prime}$ & $1.56^{\prime\prime}$ \\
Zenith Angle & 20 & 20 & 20 & 20 \\
Effective Seeing & $1.04^{\prime\prime}$ & $1.17^{\prime\prime}$ & $1.36^{\prime\prime}$ & $1.62^{\prime\prime}$ \\
\hline
\textbf{High-order Errors} & & & &  \\
\hline
Atmospheric Fitting Error & 65 & 72 & 82 & 95 \\
Bandwidth Error & 54 & 60 & 67 & 78 \\
High-order Measurement Error  & 35 & 38 & 44 & 52 \\
LGS Focal Anisoplanatism Error & 99 & 109 & 124 & 143 \\
Other High Order Errors & 64 & 65 & 68 & 72 \\
\hline
\textbf{Total High Order Wavefront Error} & \textbf{149\,nm} & \textbf{163\,nm} & \textbf{182\,nm} & \textbf{208\,nm} \\
\hline
\textbf{Tip-Tilt Errors} & & & &  \\
Tilt Bandwidth Error & $24\,$mas & $26\,$mas & $30\,$mas & $34\,$mas \\
Other Tip-Tilt Errors & $7\,$mas & $7\,$mas & $7\,$mas & $8\,$mas \\
\hline
\textbf{Total Tip/Tilt Error (one-axis)} & \textbf{25\,mas} & \textbf{27\,mas} & \textbf{31\,mas} & \textbf{35\,mas} \\
\hline
 & & & &  \\
 \hline
\textbf{Total Effective Wavefront Error} & \textbf{165\,nm} & \textbf{180\,nm} & \textbf{200\,nm} & \textbf{228\,nm} \\
\hline
 & & & &  \\
 \hline
 
 \begin{tabular}{@{}p{0.11\textwidth}p{0.075\textwidth}p{0.075\textwidth}@{}}  \textbf{Spectral Band} & $\boldsymbol\lambda$ & $\boldsymbol{\lambda/}$\textbf{D} \end{tabular} & \begin{tabular}{@{}p{0.05\textwidth}p{0.06\textwidth}@{}} \textbf{Strehl} & \textbf{FWHM} \end{tabular} & \begin{tabular}{@{}p{0.05\textwidth}p{0.06\textwidth}@{}} \textbf{Strehl} & \textbf{FWHM} \end{tabular} & \begin{tabular}{@{}p{0.05\textwidth}p{0.06\textwidth}@{}} \textbf{Strehl} & \textbf{FWHM}\end{tabular} &  \begin{tabular}{@{}p{0.05\textwidth}p{0.06\textwidth}@{}} \textbf{Strehl} & \textbf{FWHM} \end{tabular} \\
 
 \begin{tabular}{@{}p{0.11\textwidth}p{0.075\textwidth}p{0.075\textwidth}@{}} \hspace{0.045\textwidth} $r^\prime$ & $0.62\, \mu$ & $0.07^{\prime\prime}$  \end{tabular} & \begin{tabular}{@{}p{0.05\textwidth}p{0.06\textwidth}@{}} 6\% & $0.10^{\prime\prime}$  \end{tabular}  &  \begin{tabular}{@{}p{0.05\textwidth}p{0.06\textwidth}@{}}  4\% & $0.11^{\prime\prime}$    \end{tabular} &  \begin{tabular}{@{}p{0.05\textwidth}p{0.06\textwidth}@{}} 2\% & $0.14^{\prime\prime}$    \end{tabular} & \begin{tabular}{@{}p{0.05\textwidth}p{0.06\textwidth}@{}}  0\% & $0.34^{\prime\prime}$     \end{tabular} \\

  \begin{tabular}{@{}p{0.11\textwidth}p{0.075\textwidth}p{0.075\textwidth}@{}}  \hspace{0.045\textwidth}$i^\prime$ & $0.75\, \mu$ & $0.08^{\prime\prime}$   \end{tabular} & \begin{tabular}{@{}p{0.05\textwidth}p{0.06\textwidth}@{}} 14\% & $0.11^{\prime\prime}$ \end{tabular}& \begin{tabular}{@{}p{0.05\textwidth}p{0.06\textwidth}@{}}   10\% & $0.11^{\prime\prime}$   \end{tabular}  &  \begin{tabular}{@{}p{0.05\textwidth}p{0.06\textwidth}@{}}  6\% & $0.12^{\prime\prime}$ \end{tabular} &  \begin{tabular}{@{}p{0.05\textwidth}p{0.06\textwidth}@{}}   2\% & $0.15^{\prime\prime}$    \end{tabular}\\

 \begin{tabular}{@{}p{0.11\textwidth}p{0.075\textwidth}p{0.075\textwidth}@{}}  \hspace{0.045\textwidth}$z^\prime$ & $0.88\, \mu$ & $0.10^{\prime\prime}$  \end{tabular} & \begin{tabular}{@{}p{0.05\textwidth}p{0.06\textwidth}@{}}  25\% & $0.12^{\prime\prime}$ \end{tabular} &  \begin{tabular}{@{}p{0.05\textwidth}p{0.06\textwidth}@{}}  19\% & $0.12^{\prime\prime}$    \end{tabular} &  \begin{tabular}{@{}p{0.05\textwidth}p{0.06\textwidth}@{}}  13\% & $0.13^{\prime\prime}$ \end{tabular} & \begin{tabular}{@{}p{0.05\textwidth}p{0.06\textwidth}@{}}  7\% & $0.14^{\prime\prime}$      \end{tabular} \\

 \begin{tabular}{@{}p{0.11\textwidth}p{0.075\textwidth}p{0.075\textwidth}@{}}  \hspace{0.045\textwidth}J &  $1.25\, \mu$  & $0.14^{\prime\prime}$ \end{tabular} & \begin{tabular}{@{}p{0.05\textwidth}p{0.06\textwidth}@{}}49\% & $0.15^{\prime\prime}$ \end{tabular} &  \begin{tabular}{@{}p{0.05\textwidth}p{0.06\textwidth}@{}}  43\% & $0.16^{\prime\prime}$       \end{tabular} &  \begin{tabular}{@{}p{0.05\textwidth}p{0.06\textwidth}@{}}  35\% & $0.16^{\prime\prime}$ \end{tabular} & \begin{tabular}{@{}p{0.05\textwidth}p{0.06\textwidth}@{}}   26\% & $0.17^{\prime\prime}$    \end{tabular} \\

 \begin{tabular}{@{}p{0.11\textwidth}p{0.075\textwidth}p{0.075\textwidth}@{}}  \hspace{0.045\textwidth}H & $1.64\, \mu$  & $0.18^{\prime\prime}$   \end{tabular} & \begin{tabular}{@{}p{0.05\textwidth}p{0.06\textwidth}@{}} 66\% & $0.19^{\prime\prime}$ \end{tabular} &  \begin{tabular}{@{}p{0.05\textwidth}p{0.06\textwidth}@{}}  61\% & $0.20^{\prime\prime}$        \end{tabular} &  \begin{tabular}{@{}p{0.05\textwidth}p{0.06\textwidth}@{}}  54\% & $0.20^{\prime\prime}$ \end{tabular} & \begin{tabular}{@{}p{0.05\textwidth}p{0.06\textwidth}@{}}   45\% & $0.20^{\prime\prime}$     \end{tabular} \\
 
 \hline
\end{tabular}
\label{tab:error}
\end{table*}

\subsection{PSF Morphology}

\begin{figure}[h!]
    \centering
    \includegraphics[width=0.45\textwidth]{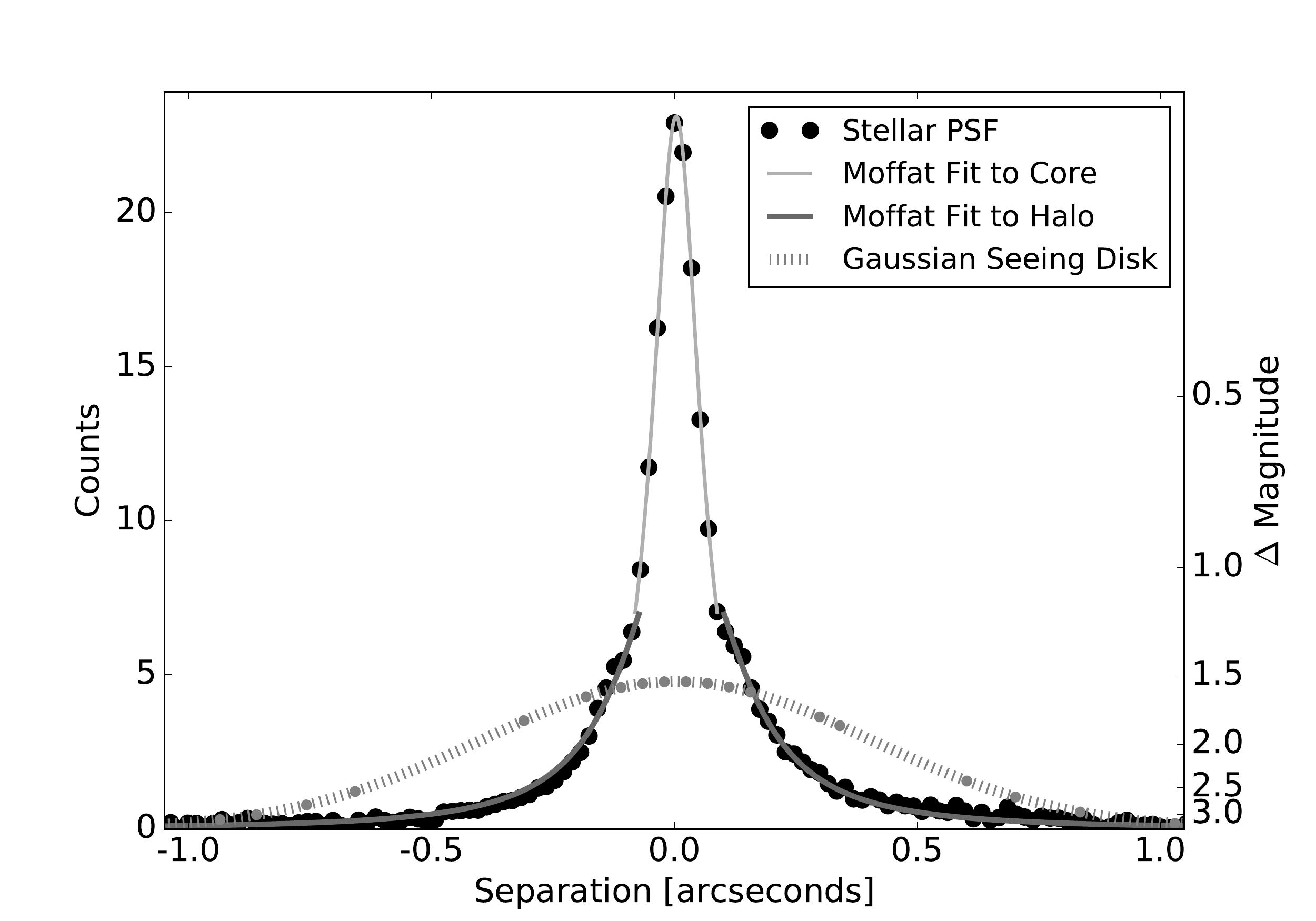}
    \caption{A 1D cut through the PSF of HIP56051 is plotted with two Moffat functions fit to the PSF core and halo, respectively. The dashed curve is a Gaussian distribution with a FWHM corresponding to the seeing measurement and an area equal to the observed PSF's area. }
    \label{fig:psf}
\end{figure}

Figure \ref{fig:psf} shows a representative Robo-AO point spread function (PSF) corresponding to the V=$10$ star HIP56051. The observation was taken in the $i^{\prime}$ band with a total exposure time of $90\,$s. The seeing at the time of the observation was $0.94^{\prime\prime}$, and the Strehl ratio of the final PSF is $10.17\%$. 

The effect of the AO system is to re-arrange the starlight from the equivalent area seeing-limited PSF (dashed curve) to the sharper, observed PSF plotted by the black points. The AO-corrected PSF includes two components: a sharp core and a broader halo, each separately fit by Moffat functions (the light and dark gray curves, respectively). The full width at half maximum (FWHM) of the Moffat function fit to the core is $0.1^{\prime\prime} \pm 0.01^{\prime\prime}$. This value is consistent with the diffraction limit of $1.028\, \lambda/D = 0.08^{\prime\prime}$.

\subsection{Contrast Curves}

\begin{figure}[h!]
    \centering
    \includegraphics[width=0.45\textwidth]{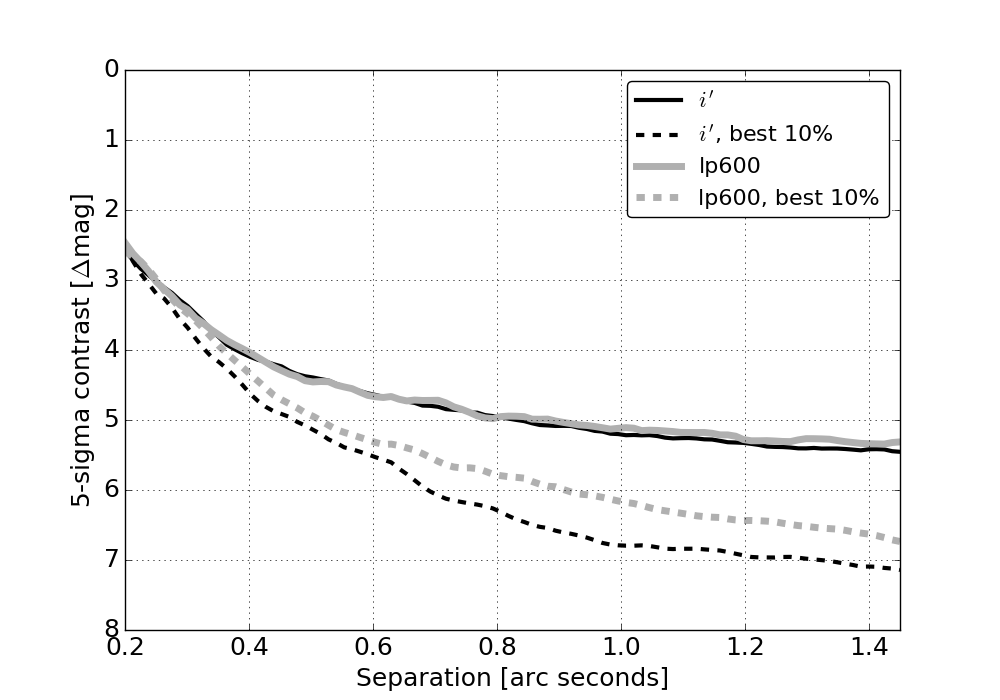}
    \caption{The contrast as a function of distance from the central star for the $i^{\prime}$ and lp600 filters. The dashed lines show the best $10\%$ contrast curves for each filter.}
    \label{fig:contrast}
\end{figure}

Section \S\ref{sec:reduction} described the ``high contrast pipeline,'' which produces $5\sigma$ contrast curves from the high pass filtered, RDI-PCA reduced science frames. Figure \ref{fig:contrast} plots the median and best $10\%$ contrast curves for $i^{\prime}$ and lp600 filter science frames. Under sub-arcsecond seeing (the best $10\%$ of cases), the contrast ratio for a $2\leq i^{\prime} \leq 16$ primary star is five and seven magnitudes at $0.5^{\prime\prime}$ and $1.0^{\prime\prime}$, respectively.

\section{Data Archive}
\label{sec:archive}
We have developed a fully automated data processing and archiving system\footnote{\url{https://github.com/dmitryduev/roboao-archive}}. The data reduction chain for an observing night proceeds as follows. At the end of each night, the visual camera data are compressed and transferred to the network storage. Next, the darks and dome flats taken at the beginning of each night are combined into master calibration files and applied to the observations. The bright star pipeline is then run on each observation followed by the computation of the Strehl ratio of the resulting image. The high contrast pipeline also produces high pass filtered, PSF-subtracted images and contrast curves for each of these processed images (see Section \ref{sec:reduction}). If the ``drizzled'' image produced by the bright star pipeline does not pass a quality check (i.e.~if a 2-component Moffat fit to the PSF has an anomalously narrow core or wide halo) then the faint star pipeline re-reduces the rapid read-out data. Additionally, the ``archiver'' processes the nightly seeing data, and generates summary plots of the seeing measurements, Strehl ratios, and contrast curves. Completing the full reduction chain for a typical night's worth of data takes a few hours.

The ``house-keeping'' system uses a  \texttt{Redis}\footnote{An efficient in-memory key-value database}-based \texttt{huey} python package\footnote{\url{https://github.com/coleifer/huey}} to manage the processing queue, which distributes the jobs to utilize all available computational resources. The processing results together with ancillary information on individual observations and system performance are stored in a \texttt{MongoDB}\footnote{\url{https://www.mongodb.com}} NoSQL database. For interactive data access, we developed a web-based interface powered by the \texttt{Flask}\footnote{\url{https://github.com/pallets/flask}} back-end. It allows the user to access previews of the processing results together with auxiliary data (e.g.~external VO images of a field), nightly summary and system performance information. The web application serves as the general interface to the database providing a sophisticated query interface and also has a number of analysis tools.

\section{Near-infrared Camera}
\label{sec:IRC}

In November 2016, we installed a NIR camera for use with Robo-AO. While similar to the camera deployed in engineering tests at the Palomar 1.5-m telescope in 2014 \citep{baranec_high-speed_2015}, the new camera uses a science-grade detector and faster readout electronics. The detector is a Mark 13 Selex ES Advanced Photodiode for High-speed Infrared Array (SAPHIRA) with an ME-911 Readout Integrated Circuit \citep{SAPHIRA1, SAPHIRA2}. It has sub-electron readnoise and  320$\times$256-pixel array with $\lambda=2.5\,\mu$m cutoff. The single-board PB1 `PizzaBox' readout electronics were developed at the Institute for Astronomy and we use 32 readout channels, each capable of a 2 Mpixel/sec sampling rate, for a maximum full-frame read rate of $\sim800$Hz.

The NIR camera attaches to the Robo-AO $f$/41 infrared camera port that accesses $\lambda > 950$nm after transmission through a dichroic. The camera has an internal cold $\lambda < 1.85\,\mu$m short-pass filter and an external warm filter wheel with J, H, clear, and blocking filters. The camera has a plate scale of $0.064^{\prime\prime}$ per pixel and field-of-view of $16.5^{\prime\prime} \times 20.6^{\prime\prime}$.  

\begin{figure}[h!]
    \centering
    \includegraphics[width=0.4\textwidth]{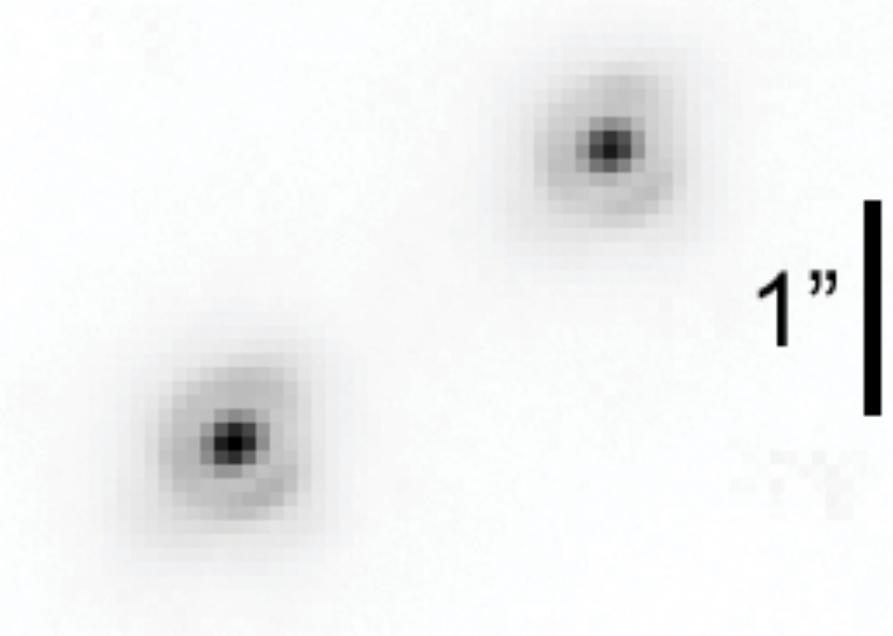}
    \caption{A $5.5\,$s image of GJ1116 taken in H-band with the near-infrared camera (linear stretch).}
    \label{fig:irimage}
\end{figure}

We achieved first light on sky in February 2017 during final testing of the upgraded TCS. Initially we used a 1 Mpixel/sec sampling rate (a full frame read rate of 390\,Hz) with detector resets every 300 reads. To create a reduced image, we first assembled difference frames between 39 consecutive reads, totaling $\sim 0.1\,$s of integration time, short enough to effectively freeze stellar image displacement. We subtracted a frame median to approximate removing the background. We then synthesized a long exposure image by registering each corrected frame on the brightest target in the field. Figure  \ref{fig:irimage} shows an example image of a binary star observed in H-band.

For the moment, data acquisition and reduction is performed manually. In the coming months, we will optimize the detector readout routines for maximum sensitivity to faint objects (including dithering for background removal), integrate the operation of the camera into the robotic queue and modify our existing data reduction pipeline to handle the NIR data. We will also investigate automating active tip-tilt correction by using either the visible or infrared camera as a tip-tilt camera, as previously demonstrated at Palomar. 

\section{Conclusion}
\label{sec:conclusion}

Robo-AO at the Kitt Peak 2.1-m telescope is the first dedicated adaptive optics observatory. Observing every clear night, Robo-AO has the capacity to undertake LGS AO surveys of large samples. For instance,  a 1000-star survey with exposure times of $60\,$s per target can be completed on the timescale of a week. 

Science programs designed to exploit Robo-AO's unique capabilities are underway. These programs include stellar multiplicity in open clusters, minor planet binarity, major planet weather variability, extragalactic object morphology, sub-stellar companions to nearby young stars, M-star multiplicity, and the influence of stellar companions on asteroseismology.
By the summer of 2017, Robo-AO will become the first LGS AO system to operate entirely autonomously, as on-going upgrades to the 2.1-m telescope will remove the need for a human observer.

\acknowledgments
The Robo-AO team thanks NSF and NOAO for making   the Kitt Peak 2.1-m telescope available. We thank the observatory staff at Kitt Peak for their efforts to assist Robo-AO KP operations.  Robo-AO KP is a partnership between the California Institute of Technology, the University of Hawai`i, the University of North Carolina at Chapel Hill, the Inter-University Centre for Astronomy and Astrophysics (IUCAA) at Pune, India, and the National Central University, Taiwan. The Murty family feels very happy to have added a small value to this important project. Robo-AO KP is also supported by grants from the John Templeton Foundation and the Mt.~Cuba Astronomical Foundation. The Robo-AO instrument was developed with support from the National Science Foundation under grants AST-0906060, AST-0960343, and AST-1207891, IUCAA, the Mt.~Cuba Astronomical Foundation, and by a gift from Samuel Oschin. These data are based on observations at Kitt Peak National Observatory, National Optical Astronomy Observatory (NOAO Prop. ID: 15B-3001), which is operated by the Association of Universities for Research in Astronomy (AURA) under cooperative agreement with the National Science Foundation. C.B. acknowledges support from the Alfred P. Sloan Foundation.

\facility{KPNO:2.1m (Robo-AO)}

\bibliographystyle{yahapj}
\bibliography{references.bib}

\appendix
\section{Telescope Jitter}
\label{apend:jitter}
\begin{figure*}[htp]
  \centering
  \subfigure[Kitt Peak mean subtracted RA centroids]{\includegraphics[width=0.45\textwidth]{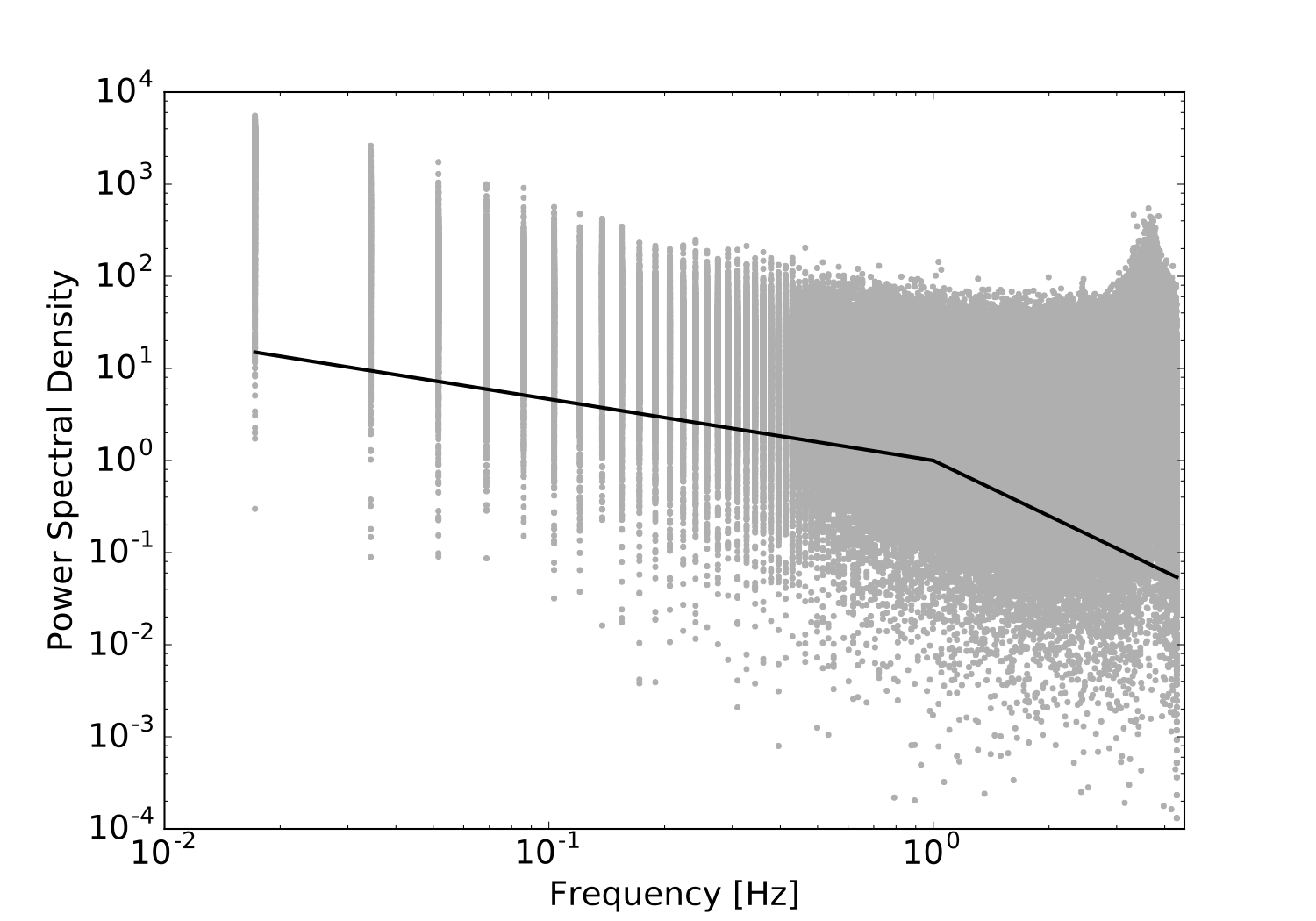}}\quad
  \subfigure[Palomar mean subtracted RA centroids]{\includegraphics[width=0.45\textwidth]{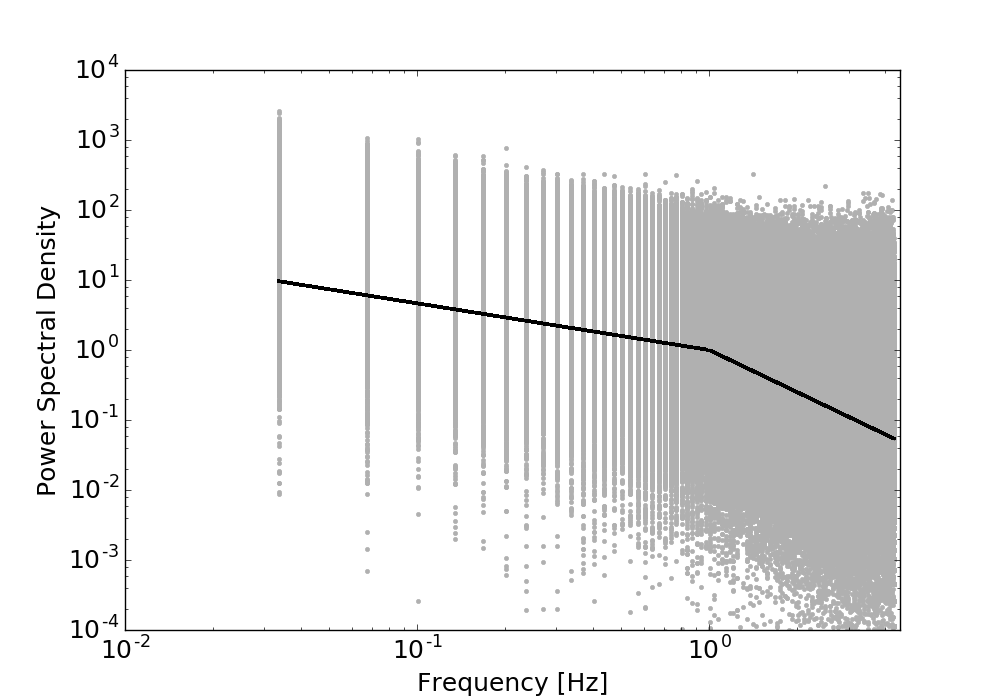}}
\caption{The power spectral densities of the mean subtracted RA target positions for each sub-exposure at Kitt Peak (a) and Palomar (b). The peak at $\sim 3.7\,$Hz is present at Kitt Peak, but not at Palomar. The solid black lines show the theoretical power-law dependencies of the tilt: $f^{-2/3}$ at low frequencies, and $f^{-2}$ for $1-10\,$Hz \citep{hardy_adaptive_1998}.}
\label{fig:psds}
\end{figure*}

\begin{figure}[h!]
    \centering
    \includegraphics[width=0.45\textwidth]{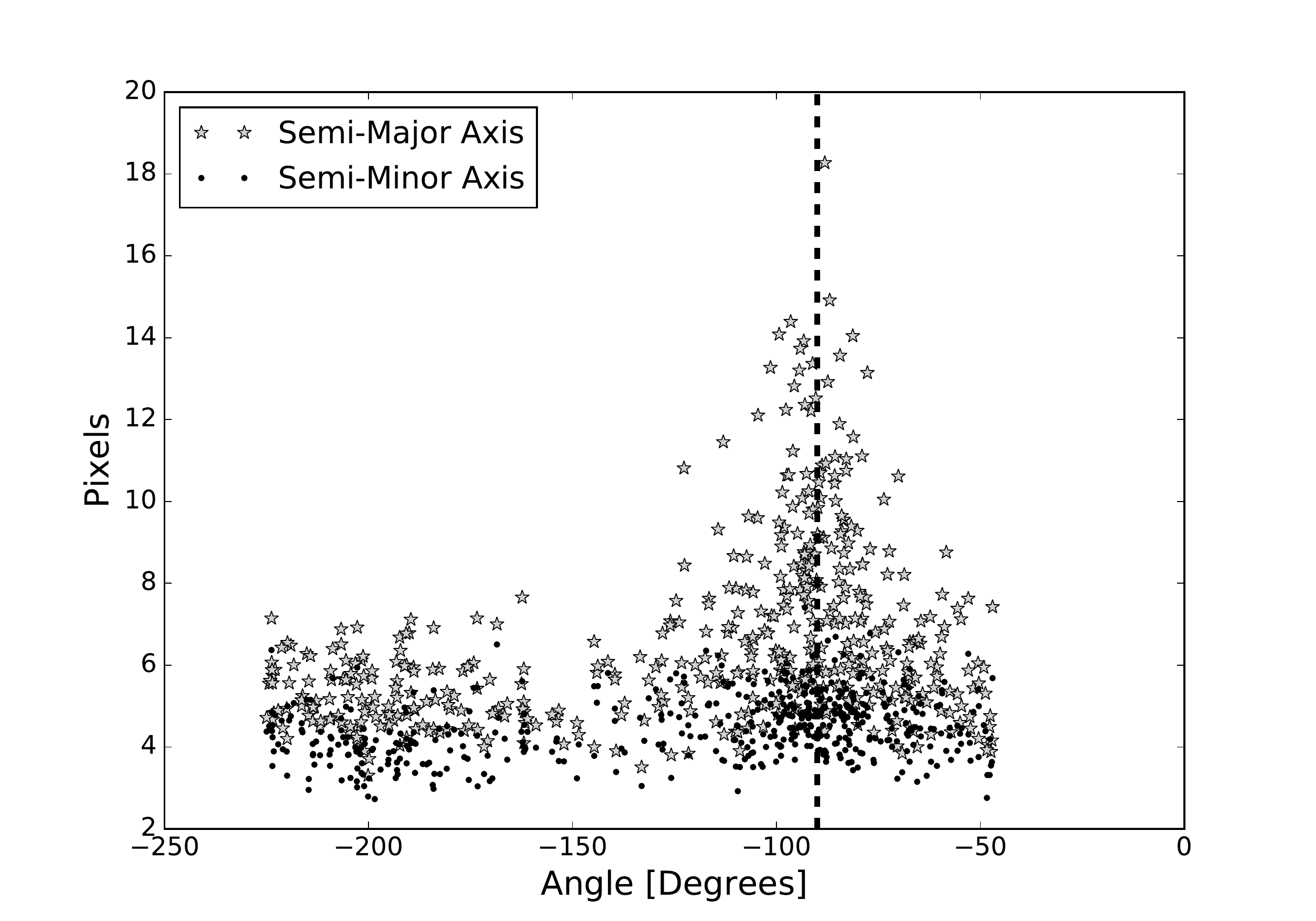}
    \caption{For a test observation, the standard deviation along the semi-major and semi-minor axes of 2D Gaussian fits to each 0.116s sub-exposure are plotted versus the rotation angle of the Gaussian. Here, $-90^{\circ}$ (dashed black line) indicates that the semi-major axis lies along the RA-axis. Clearly, the PSF is elongated along the RA-axis. }
    \label{fig:smearing}
\end{figure}

After moving Robo-AO from the Palomar 1.5-m telescope to the Kitt Peak 2.1-m telescope, the median Strehl ratio across all wavelengths was initially reduced from $5.8\%$ to $3.2\%$. The source of this degradation was a $\sim3.7\,$Hz vibration in the RA axis. Because Robo-AO mitigates tip/tilt by post facto shift and add rather than a real-time loop, and because its framerate is typically only $8.6\,$Hz, the targets were smeared in the RA direction. Figure \ref{fig:psds} a and b show the power spectral densities of the mean subtracted RA centroid positions of targets observed at Kitt Peak and Palomar, respectively. The peak at $\sim3.7\,$Hz is clear in the Kitt Peak data, but is not present at Palomar. The RA-axis smearing for a single test observation is demonstrated in Figure \ref{fig:smearing}. 

\begin{figure}[h!]
    \centering
    \includegraphics[width=0.45\textwidth]{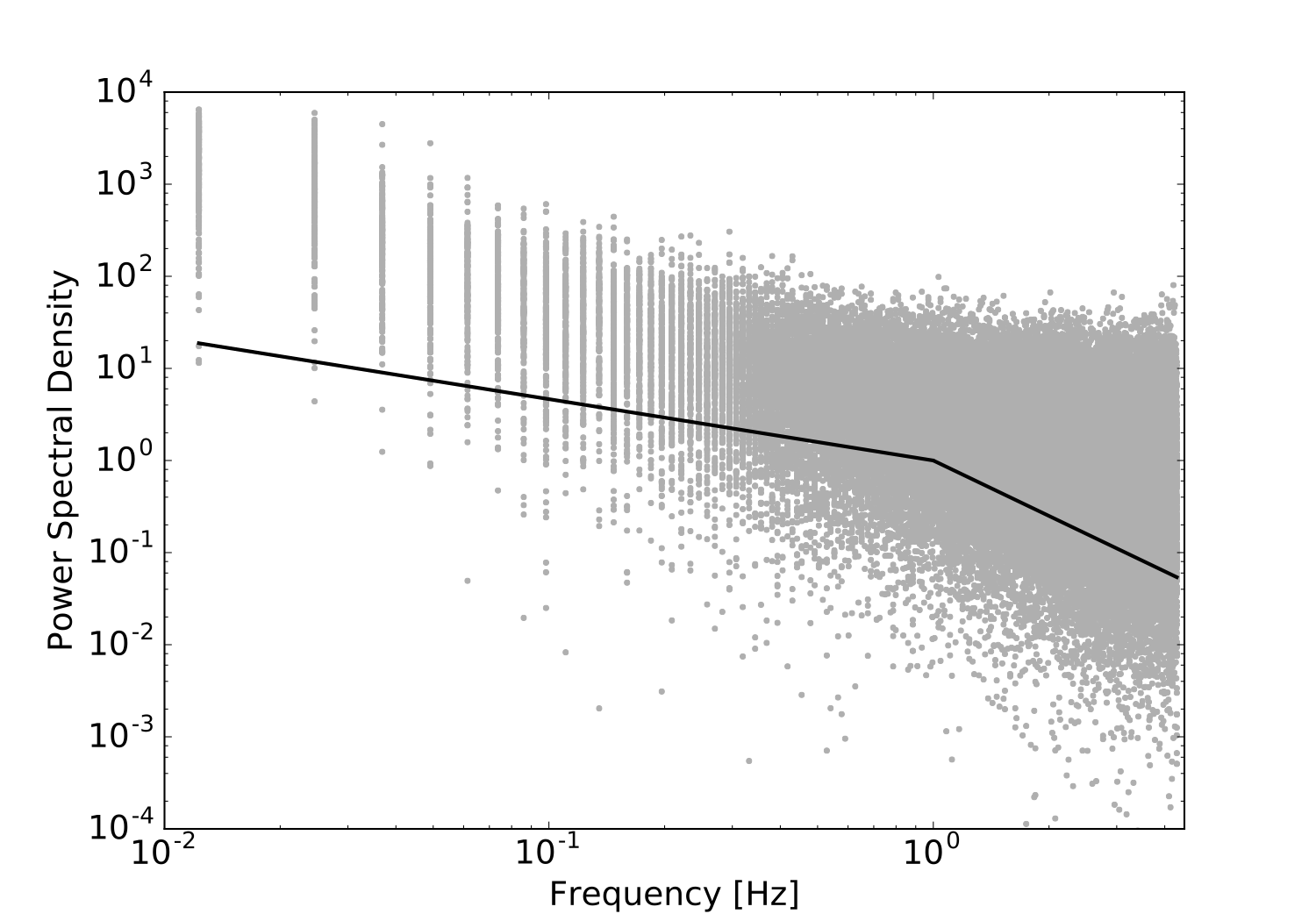}
    \caption{The power spectral densities of the mean subtracted RA target positions for the Kitt Peak sub-exposures since the telescope control upgrade (22 February 2017 through 8 March 2017). The peak that was present in Figure \ref{fig:psds}a is eliminated. }
    \label{fig:psd_improved}
\end{figure}

\begin{figure}[h!]
    \centering
    \includegraphics[width=0.45\textwidth]{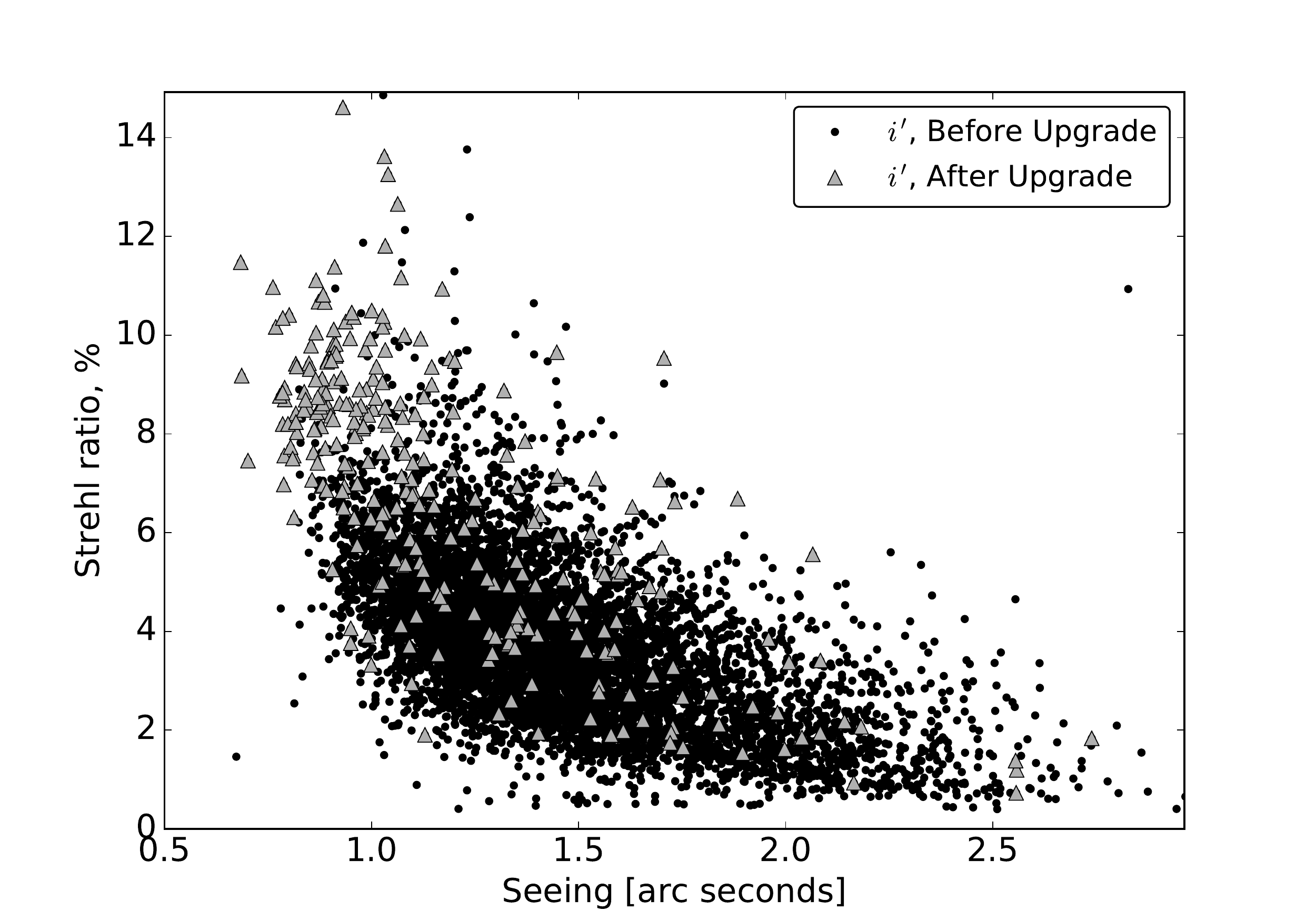}
    \caption{Strehl ratios of the observations taken in $i^{\prime}$ as a function of the seeing scaled to 500 nm before (December 2015 through 22 February 2017; black points) and after (22 February 2017 through 10 March 2017; gray stars) the enhancements. Note the significant improvement for seeing under $\approx 1.1$ arcseconds.}
    \label{fig:strehl_comparison}
\end{figure}

The jitter was mitigated by two changes to the system. First, the KPNO staff noticed a ticking sound corresponding to each rotation of the telescope drive worm gear, which was solved by lubrication. This step reduced the height of, but did not eliminate, the PSD peak. Second, we took a test observation in which only sidereal tracking was enabled, and all fine computer guiding was turned off. The peak was absent in this test observation, leading us to conclude that the telescope control system (TCS) was giving erroneous commands that ``kicked'' the telescope's position. The TCS was replaced in the winter of 2017 to allow the Robo-AO robotic system to fully control the telescope's motion, eliminating the remaining RA jitter (Figure \ref{fig:psd_improved}). Figure \ref{fig:strehl_comparison} shows a comparison of the Strehl ratios versus the seeing before and after the TCS upgrade. 

\end{document}